\documentclass[aps,prd, preprintnumbers,showpacs,nofootinbib, twocolumn]{revtex4}
\usepackage{graphicx}
\usepackage{amsmath}
\usepackage{bm}
\usepackage{color}

\def\be{\begin{equation}}
\def\ee{\end{equation}}
\def\bea{\begin{eqnarray}}
\def\eea{\end{eqnarray}}

\begin{document}

\title{Isotropic stars in general relativity }
\author{M. K. Mak}
\email{mkmak@vtc.edu.hk}
\affiliation{Department of Computing and Information Management, Hong Kong
Institute of Vocational Education, Chai Wan, Hong Kong, P. R. China}
\author{T. Harko}
\email{tiberiu.harko@gmail.com}
\affiliation{Department of Mathematics, University College London, Gower Street, London WC1E 6BT, United Kingdom}

\begin{abstract}
We present a general solution of the Einstein gravitational field equations  for the static spherically symmetric gravitational  interior spacetime of an isotropic fluid sphere. The solution is obtained by transforming the pressure isotropy condition, a second order ordinary differential equation, into a Riccati type first order differential equation, and using a general integrability condition for the Riccati equation. This allows us to obtain an exact non-singular solution of the interior field equations for a fluid sphere, expressed in the form of infinite power series.  The physical features of the solution are studied in detail numerically by cutting the infinite series expansions, and restricting our numerical analysis by taking into account only $n=21$ terms in the power series representations of the relevant astrophysical parameters.   In the present model all physical quantities (density, pressure, speed of sound etc.) are finite at the center of the sphere. The physical behavior of the solution essentially depends on the equation of state of the dense matter at the center of the star. The stability properties of the model are also analyzed in detail for a number of central equations of state, and it is shown that it is stable with respect to the radial adiabatic perturbations. The astrophysical analysis indicates that this solution can be used as a realistic model for static general relativistic high density objects, like neutron stars.
\end{abstract}

\maketitle

\section{ Introduction}

Spherically symmetric problems are very important in general relativity
because phenomena such as black holes, neutron stars, quark stars and
gravitational collapse have been found in the class of system with spherical
symmetry. Relativistic stellar models have been studied ever since the first
solution of Einstein's field equation for the interior of a compact object
in hydrostatic equilibrium was obtained by Karl Schwarzschild in 1916 \cite{Sch}. The search
for the exact solutions describing static isotropic and anisotropic stellar
type configurations has continuously attracted the interest of physicists. An important moment in the development of this field is the paper by Tolman \cite{Tol}, in which a method to obtain explicit analytical solutions of the field equations was proposed. Instead of using the equation of state of matter to close the system of the field equations, Tolman  proposed  to introduce an additional equation necessary to give a determinate problem in the form of some
{\it ad hoc} relation between the metric tensor components, as to make the resulting set of field equations easy to solve. Based on this methodology eight solutions of the field equations were obtained in \cite{Tol}, and this approach still continues to be one of the important methods in obtaining exact interior solutions of the gravitational field equations for fluid spheres. Buchdahl \cite{Buch} made an important contribution to the study of the fluid spheres by obtaining the famous bound on the mass radius ratio for stable general relativistic spheres, $2GM/c^2R\leq 8/9$. An exact non-singular solution, based on a particular choice of the mean density inside the star, was also obtained.

There are very few exact interior solutions (both isotropic and anisotropic)
of the gravitational field equations satisfying the required general
physical conditions inside the star. From $127$ published solutions analyzed
by Delgaty and Lake \cite{Del} only $16$ satisfy all the conditions. But the
study of the interior of general relativistic stars via finding exact
solutions of the field equations is still an active field of research. The spacetime metrics corresponding to the class of all
static spherically symmetric perfect fluid geometries were explicitly characterized in \cite{Rah}. The results may be useful whenever there is some uncertainty regarding the actual equation of state of the dense matter inside the general relativistic star.  Further generalizations of the Buchdahl bound by placing a number of constraints on the interior geometry (the metric components), on the local acceleration due to gravity, on various combinations of the internal density and pressure profiles and on the internal compactness $2m(r)/r$ of static fluid spheres were obtained in \cite{Mar}. An algorithm based on the choice of a single monotone function (subject to boundary conditions),  which generates all regular static spherically symmetric perfect-fluid solutions of Einstein's equations was presented in \cite{L0}. With the help of this algorithm an infinite number of previously unknown physically interesting exact solutions can be constructed. A variant of the algorithm developed in \cite{L0} was introduced in \cite{Mar1} by recasting it in terms of variables with a clear physical meaning, like the average density and the locally measured acceleration due to gravity. The formalism can be used to understand the relationships among some known exact solutions, and generate several new exact solutions of the Einstein field equations for the static fluid sphere.

Several transformation theorems that map perfect fluid spheres into perfect fluid spheres were developed in \cite{V1}. The transformation theorems can be used to develop a systematic way of classifying the set of all perfect fluid spheres, and they also led to unexpected connections between previously known perfect fluid spheres, and to previously solutions of the field equations describing perfect fluid spheres.  "Solution generating" theorems for the Tolman-Oppenheimer-Volkov equation, whereby any given solution can be "deformed" to a new solution, were developed in \cite{V2}. The theorems work directly in terms of the physical observables of the fluid sphere - the pressure profile and the density profile, respectively. The "deformed" solutions of the Tolman-Oppenheimer-Volkov equation are conveniently parameterized in terms of $\delta \rho_c$ and $\delta p_c$, the finite shifts in the central density and central pressure, respectively. Several new and significant transformation theorems that
map perfect fluid spheres to perfect fluid spheres using both usual and
unusual coordinate systems, such as Schwarzschild (curvature), isotropic,
Gaussian polar (proper radius), Synge isothermal (tortoise) and Buchdahl
coordinates were discussed in \cite{V3}.

The Einstein static universe was transformed
into a class of physically acceptable static fluid spheres, whose physical
properties were written down in an explicit form in \cite{L1}. An explicit
four-fold infinity of physically acceptable exact perfect fluid solutions of
Einstein's equations by way of conformal transformations of physically
unacceptable solutions was discussed in \cite{L2}. Using these transformations as a basis,
one can classify different types of perfect fluid sphere solutions.

The necessary and sufficient condition for universality of the
Schwarzschild interior solution describing a uniform density sphere, for all $n\geq
4 $ dimensions, in the framework of Einstein-Gauss-Bonnet gravity, was presented in \cite{Dad}.
A standard polynomial expansion
technique was used to show the existence of a relation between
polytropic model and the description of gas spheres at finite
 temperature in \cite{Souza}.

The study of general relativistic compact objects is of fundamental
importance for astrophysics. After the discovery of pulsars and explanation
of their properties by assuming them to be rotating neutron stars, the
theoretical investigation of superdense stars has been done using both
numerical and analytical methods and the parameters of neutron stars have
been worked out by general relativistic treatment.

It is the purpose of this paper to introduce an exact solution of the Einstein gravitational field equations, describing the interior of a general relativistic isotropic fluid sphere. The condition of the pressure isotropy can be formulated as a second order homogenous equation, which in turn can be transformed into a first order Riccati type differential equation. The exact solution of the Riccati equation can be found by using an integrability condition for this equation, thus allowing the construction of an exact solution of the gravitational field equations, without the use of any {\it ad hoc} assumptions on the physical or geometrical quantities. The obtained solution is non-singular at the center of the fluid sphere, and it is expressed in the form of infinite power series. The physical properties of the corresponding stellar model are analyzed in detail numerically,  by cutting the infinite series expansions, and restricting our numerical analysis by taking into account only $n=21$ terms in the power series representations of the relevant astrophysical parameters.  The solution presented in this paper satisfies all the physical conditions required for realistic stellar models, like monotonicity of the density and pressure, a speed of some smaller than the speed of light for most of the equations of state of the matter at the center of the star, and matching on the vacuum boundary with the Schwarzschild exterior solution. The stability properties of the model have been analyzed in detail, and we have shown that the model is stable with respect to radial adiabatic perturbations. From an astrophysical point of view the solution describes an ultra-compact massive stellar type object, with radius of around 10 km, and a mass of the order of $2.1M_{\odot}$, corresponding to a central density of $3\times 10^{15}$ g/cm$^3$.

The present paper is organized as follows. The gravitational field equations and the integrability condition of the Riccati equation are presented in Section~\ref{sect2}. The general solution of the field equations is obtained in Section~\ref{sect3}. The physical properties of the corresponding stellar model are analyzed in Section~\ref{sect4}. We discuss and conclude our results in Section~\ref{sect5}.

\section{Geometry, field equations, and the integrability condition}\label{sect2}

In standard coordinates $x^{i}=\left( t,r,\theta ,\phi \right) $, the
 line element for a static spherically symmetric space-time takes the
form
\be  \label{41}
ds^{2}=A^{2}(r)c^2dt^{2}-B^{-1}(r)dr^{2}-
r^{2}\left( d\theta ^{2}+\sin ^{2}\theta d\phi ^{2}\right).
\ee

Einstein's gravitational field equations are
\begin{equation}  \label{5}
R_{i}^{k}-\frac{1}{2}R\delta _{i}^{k}=\frac{8\pi G}{c^4}T_{i}^{k}.
\end{equation}

For an isotropic spherically symmetric matter distribution the components of
the energy-momentum tensor are of the form \cite{LaLi}
\begin{equation}  \label{6}
T_{i}^{k}=\left( \rho c^2+p\right) u_{i}u^{k}-p\delta _{i}^{k},
\end{equation}
where $u^{i}$ is the four-velocity $u^{i}=\delta _{0}^{i}$, $\rho $ is the
energy density, and $p$ is the pressure, respectively. For any physically
acceptable stellar models, we require the condition that the energy density and the pressure
are positive and finite at all points inside the fluid spheres.

For the metric given by Eq.~(\ref{41}) the gravitational field equations
that follows from Eqs.~(\ref{5}) are
\begin{equation}  \label{7_1}
\frac{8\pi G}{c^2}\rho =\frac{1-B}{r^{2}}-\frac{1}{r}\frac{dB}{dr},
\end{equation}
\begin{equation}
\frac{8\pi G}{c^4}p=2\frac{1}{A}\frac{dA}{dr}\frac{B}{r}+\frac{B-1}{r^{2}},
\label{7}
\end{equation}
and
\be\label{8}
\left( \frac{1}{A}\frac{dA}{dr}+\frac{1}{r}\right)\frac{dB}{dr} +
2B\left( \frac{%
1}{A}\frac{d^2A}{dr^2}-\frac{1}{rA}\frac{dA}{dr}-\frac{1}{r^{2}}\right) +\frac{2%
}{r^{2}}=0 ,
\ee
respectively. From a physical point of view Eq.~(\ref{8}) represents
the condition of the pressure isotropy inside the star. Eq.~(\ref{7_1}) can
be immediately integrated to give
\begin{equation}
B(r)=1-\frac{2Gm(r)}{c^2r},
\end{equation}
where $m(r)=4\pi \int_{0}^{r} {\rho \left( \zeta \right) \zeta ^2d\zeta }$ represents the total mass content of the distribution within
the fluid sphere of radius $r$. In the following we introduce the notations
\begin{equation}
x =r^{2},
\end{equation}
\begin{equation}
\eta \left( r\right) =\frac{Gm(r)}{c^2r^{3}},
\end{equation}
and
\begin{equation}
B(x)=1-2x \eta (x),
\end{equation}
respectively. At the center of the star the function $\eta $ has the value
\begin{equation}
\lim_{r\longrightarrow 0}\eta (r)=\frac{ 4\pi G}{3c^2} \rho _{c},
\end{equation}
where $\rho _{c}$ is the central density. By taking into account the
definition of $m$ as $dm/dr=4\pi \rho r^2$ and of $\eta $, we can express
the density $\rho $ as
\begin{equation}
\frac{4\pi G}{c^2}\rho (r)=r\frac{d\eta }{dr}+3\eta ,
\end{equation}
or, equivalently,
\begin{equation}
\frac{4\pi G}{c^2}\rho (x)=2x\frac{d\eta (x)}{dx}+3\eta (x).
\end{equation}

Hence, we can express the condition of the isotropy of the pressure, given by Eq.~(\ref{8}), in the form \cite{Buch}
\begin{equation}
\left( 1-2x\eta \right) A^{^{\prime \prime }}-\left( x\eta ^{\prime }+\eta
\right) A^{\prime }-\frac{1}{2}\eta ^{\prime }A=0,
\end{equation}%
or, equivalently,
\begin{equation}
\frac{A^{\prime \prime }}{A}-\frac{x\eta ^{\prime }+\eta }{1-2x\eta }\frac{%
A^{\prime }}{A}-\frac{1}{2}\frac{\eta ^{\prime }}{1-2x\eta }=0,  \label{1}
\end{equation}
where $'=d/dx$.

By denoting $u=A^{\prime }/A$, we have $A^{\prime \prime }/A=u^{\prime
}+u^{2}$, and Eq. (\ref{1}) can be written as
\begin{equation}
u^{\prime }=\frac{1}{2}\frac{\eta ^{\prime }}{1-2x\eta }+\frac{x\eta
^{\prime }+\eta }{1-2x\eta }u-u^{2}.  \label{2a}
\end{equation}

At this moment it is useful to observe that $x\eta ^{\prime }+\eta =d\left(
x\eta \right)/dx$, and denote
\begin{equation}
v=x\eta .
\end{equation}

Then $\eta +x\eta ^{\prime }=v^{\prime }$, and $\eta ^{\prime }=v^{\prime
}/x-v/x^{2}$, respectively. Hence Eq.~(\ref{2a}) becomes
\begin{equation}
u^{\prime }=\frac{1}{2\left( 1-2v\right) }\left( \frac{v^{\prime }}{x}-\frac{%
v}{x^{2}}\right) +\frac{v^{\prime }}{1-2v}u-u^{2}.  \label{3}
\end{equation}

\subsection{ The integrability condition for the Riccati equation}

Eq.~(\ref{3}) is a Riccati type ordinary differential equation of the form
\begin{equation}
\frac{dy}{dx}=P(x)+Q(x)y+R(x)y^{2},  \label{ricc1}
\end{equation}%
where $P$, $Q$, $R$ are arbitrary real functions of $x$, with $P,Q,R\in
C^{\infty }(I)$, defined on a real interval $I\subseteq \Re $. A general
integrability condition of the Riccati equation can be obtained as follows.
From an algebraic point of view Eq.~(\ref{ricc1}) is a quadratic equation in
$y$. We consider that its particular solutions $y_{\pm }^{p}(x)$ can be
represented as
\begin{equation}
y_{\pm }^{p}(x)=\frac{-Q(x)\pm \sqrt{f(x)}}{2R\left( x\right) },  \label{y0}
\end{equation}%
where we have introduced the new function $f(x)\in C^{\infty }(I)$, defined
as
\begin{equation}
f\left( x\right) =Q^{2}\left( x\right) -4R\left( x\right) \left[ P\left(
x\right) -\frac{dy}{dx}\right] .
\end{equation}

The requirement that the functions $y_{\pm }^{p}(x)$ given by Eq.~(\ref{y0}) are particular
solutions of the Riccati Eq.~(\ref{ricc1}), restricts the expression of $P(x)$
to the form
\begin{equation}
P(x)=\frac{d}{dx}\left[ \frac{-Q(x)\pm \sqrt{f(x)}}{2R(x)}\right] +\frac{%
Q^{2}(x)-f(x)}{4R(x)}.  \label{4_1}
\end{equation}

By substituting $P(x)$ given by Eq.~(\ref{4_1}) into Eq.~(\ref{ricc1}), we
obtain a Riccati equation having the form
\begin{eqnarray}
\frac{dy}{dx} &=&\frac{d}{dx}\left[ \frac{-Q(x)\pm \sqrt{f(x)}}{2R(x)}\right]
+  \notag  \label{2} \\
&&\frac{Q^{2}(x)-f(x)}{4R(x)}+Q(x)y+R(x)y^{2},  \notag \\
&&
\end{eqnarray}%
where $f(x)$ is a solution generating function to the auxiliary Riccati Eq.~(%
\ref{2}). Therefore we obtain the following

\textbf{Theorem.} If the coefficients $P(x)$, $Q(x)$ and $R(x)$ of a Riccati
differential equation satisfy the condition (\ref{4_1}), with $f(x)\in
C^{\infty }(I)$ an arbitrary function defined on a real interval $I\subseteq
\Re $, then the general solution of the Riccati Eq.~(\ref{2}) is represented
by
\begin{eqnarray}\label{6a}
y_{\pm }(x) &=&\frac{e^{\pm \int \sqrt{f(x)}dx}}{C_{\pm}-\int R(x)e^{\pm \int
\sqrt{f(x)}dx}dx}+  \nonumber\\
&&\left[ \frac{-Q(x)\pm \sqrt{f(x)}}{2R(x)}\right] ,
\end{eqnarray}
where $C_{\pm}$ are  arbitrary integration constants \cite{Ma}.

In the particular case $f(x)\equiv 0$, the integrability condition of the
Riccati equation becomes
\begin{equation}
P(x)=-\frac{d}{dx}\frac{Q(x)}{2R(x)}+\frac{Q^{2}(x)}{4R(x)},  \label{int2}
\end{equation}%
and the general solution of the Riccati equation satisfying the
integrability condition given by Eq.~(\ref{int2}) is obtained as
\begin{equation}
y(x)=-\frac{Q(x)}{2R(x)}+\frac{1}{C-\int R(x)dx},
\end{equation}
where $C$ is an arbitrary integration constant.

\section{The general solution of the field equations satisfying the Riccati
equation integrability condition}\label{sect3}

The basic equation describing the physical and geometrical
properties of the interior of the general relativistic fluid spheres is given by the Riccati
Eq.~(\ref{3}). The coefficients of this equation are represented by
\begin{equation}
P=\frac{1}{2\left( 1-2v\right) }\left( \frac{v^{\prime }}{x}-\frac{v}{x^{2}}%
\right) ,
\end{equation}%
\begin{equation}
Q=\frac{v^{\prime }}{1-2v}=-\frac{1}{2}\frac{d}{dx}\ln \left( 1-2v\right) ,
\end{equation}%
and
\begin{equation}
R=-1,
\end{equation}%
respectively. In the following we restrict our study to the case $f(x)\equiv
0$ only. Therefore the integrability condition of Eq.~(\ref{3}), given by
Eq.~(\ref{int2}), becomes
\be
\frac{1}{\left( 1-2v\right) }\left( \frac{v^{\prime }}{x}-\frac{v}{x^{2}}%
\right) =  \label{4}
\frac{d}{dx}\frac{v^{\prime }}{\left( 1-2v\right) }-\frac{v^{\prime 2}}{%
2\left( 1-2v\right) ^{2}}.
\ee

By denoting
\begin{equation}
w=1-2v, v=\frac{1-w}{2},
\end{equation}
Eq.~(\ref{4}) can be written as
\begin{equation}
w^{\prime \prime }-\frac{3}{4}\frac{w^{\prime 2}}{w}-\frac{w^{\prime }}{x}+%
\frac{w}{x^{2}}-\frac{1}{x^{2}}=0.  \label{10}
\end{equation}

By introducing a new dependent variable $V$ so that
\begin{equation}
w=1-2v=V^{4},
\end{equation}
Eq.~(\ref{10}) takes the form
\begin{equation}
4x^{2}V^{\prime \prime }-4xV^{\prime }+V=\frac{1}{V^{3}}.  \label{13}
\end{equation}

In terms of $V$ the function $\eta $ is given by
\begin{equation}
\eta (x)=\frac{Gm(x)}{c^{2}x^{3/2}}=\frac{v}{x}=\frac{1-V^{4}}{2x}.
\label{20}
\end{equation}

\subsection{The solution of the integrability condition}

The integrability condition of the gravitational field equations, given by
the second order non-linear differential Eq.~(\ref{13}), has the exact particular
solution $V=1$. In order to obtain a general solution for $V$, we consider
that $V(x)$ can be represented as a power series of the form
\begin{equation}
V(x)=1-a_{1}x+\sum_{n=2}^{\infty }a_{n}x^{n},  \label{34}
\end{equation}%
where $a_{n}\in \Re $. Next we expand $1/V^{3}(x)$ in power series near $x=0$%
, so that
\be
\frac{1}{\left( 1-a_{1}x+\sum_{n=2}^{\infty }a_{n}x^{n}\right) ^{3}}%
=1-a_{1}x+
\sum_{n=2}^{\infty }b_{n}\left( a_{1},a_{2},...,a_{n}\right) x^{n},
\label{35}
\ee%
where the coefficients $b_{n}$ are polynomial functions of $a_{n}$. The
first few terms of the series expansion are
\bea
&&\frac{1}{\left( 1-a_{1}x+a_{2}x^{2}+a_{3}x^{3}+...\right) ^{3}}=
1+3a_{1}x+\nonumber\\
&&3\left( 2a_{1}^{2}-a_{2}\right) x^{2}+
\left( 10a_{1}^{3}-12a_{1}a_{2}-3a_{3}\right) x^{3}+\nonumber\\
&&3\left(5a_{1}^{4}-10a_{1}^{2}a_{2}-4a_{1}a_{3}+2a_{2}^{2}-a_{4}\right)x^4+...
\end{eqnarray}

After substituting Eqs.~(\ref{34}) and (\ref{35}) in Eq.~(\ref{13}) it turns
out that these series expansions solve \textit{exactly} the equation for all
$n$, $n=1,2,...,\infty$. The first few terms obtained after the substitution
of the series expansions in Eq.~(\ref{13}) are
\bea
&&2\left(3a_1^2-2a_2\right)x^2+
2\left(5a_1^3-6a_1a_2-8a_3\right)x^3+3\Big(5a_{1}^{4}-\nonumber\\
&&10a_{1}^{2}a_{2}-4a_{1}a_{3}+2a_{2}^{2}-12a_{4}\Big)x^4+....=0.
\eea

Moreover, all coefficients $a_{n}$, $n\neq 1$, can be obtained as a function
of $a_{1}$, so that the \textit{exact} solution of Eq.~(\ref{13}) can be
obtained in the form of an infinite power series as
\begin{eqnarray}
V(x) &=&1-a_{1}x+\frac{3\,a_{1}^{2}}{2}x^{2}-\frac{a_{1}^{3}}{2}x^{3}-\frac{%
7\,a_{1}^{4}}{24}x^{4}+
\frac{a_{1}^{5}}{8}x^{5}+\nonumber\\
&&\frac{111\,a_{1}^{6}}{400}x^{6}+\frac{%
23\,a_{1}^{7}}{3600}x^{7}-
\frac{46117\,a_{1}^{8}}{156800}x^{8}-\nonumber\\
&&\frac{65273\,a_{1}^{9}}{470400}x^{9}+
\frac{4624547\,a_{1}^{10}}{15240960}x^{10}+\nonumber\\
&&\frac{28597727\,a_{1}^{11}}{%
90720000}x^{11}-\frac{83015551249\,a_{1}^{12}}{307359360000}x^{12}+...\nonumber\\
\end{eqnarray}

To find out the physical interpretation of $a_{1}$, we substitute $V$ given
by Eq.~(\ref{34}) into Eq.~(\ref{20}) for the mean density $\eta $. Using
the l'Hospital rule, we obtain the limit of mean density $\eta $ for $x$
approaching zero given by
\bea
&&\lim_{x\rightarrow 0}\eta (x)=\lim_{x\rightarrow 0}\frac{1-V^{4}(x)}{2x}%
=\nonumber\\
&&\lim_{x\rightarrow 0}\frac{1-\left[ 1-a_{1}x+\sum_{n=2}^{\infty }a_{n}x^{n}%
\right] ^{4}}{2x}=\nonumber\\
&&\lim_{x\rightarrow 0}\frac{\frac{d}{dx}\left[ 1-\left[
1-a_{1}x+\sum_{n=2}^{\infty }a_{n}x^{n}\right] ^{4}\right] }{2}=\nonumber\\
&&
\lim_{x\rightarrow 0}\Bigg\{ -2\left( 1-a_{1}x+\sum_{n=2}^{\infty }a_{n}x^{n}%
\right) ^{3}\times \nonumber\\
&&\left( -a_{1}+\sum_{n=2}^{\infty }a_{n}nx^{n-1}\right) \Bigg\}.
\eea

Therefore we obtain
\begin{equation}
\lim_{x\rightarrow 0}\eta (x)=4\pi \frac{G\rho _{c}}{3c^{2}}=2a_{1},
\end{equation}%
giving
\begin{equation}
2a_{1}=\frac{4\pi G\rho _{c}}{3c^{2}}=\frac{1}{x_{R}}=\frac{1}{R_{c}^{2}},
\end{equation}%
or
\be
R_{c} =\sqrt{\frac{3c^{2}}{4\pi G\rho _{c}}}=
1.036\times 10^{6}\times
\left( \frac{\rho _{c}}{3\times
10^{15}\mathrm{%
g/cm^{3}}}\right) ^{-1/2}\;\mathrm{cm}.
\ee

By denoting $\xi =x/x_{R}$, $V$ can be represented as
\begin{eqnarray}
V\left( \xi \right) &=&1-\frac{1}{2}\xi +\frac{3}{8}\xi ^{2}-\frac{1}{16}\xi
^{3}-\frac{7}{384}\xi ^{4}+
\frac{1}{256}\xi ^{5}+\nonumber\\
&&\frac{111}{25600}\xi ^{6}+\frac{23}{460800}\xi ^{7}-
\frac{46117}{40140800}\xi ^{8}-\nonumber\\
&&\frac{65273}{240844800}\xi ^{9}+
\frac{4624547}{15606743040}\xi ^{10}+ \nonumber\\
&&\frac{28597727}{185794560000}\xi
^{11}-
\frac{83015551249}{1258943938560000}\xi ^{12}+....\nonumber\\
\end{eqnarray}

As a function of $\xi $, $\eta $ can be written as
\begin{equation}
\eta (\xi )=\frac{4\pi G}{3c^2}\rho _c\frac{1-V^4(\xi )}{2\xi }.
\end{equation}

\subsection{The general solution of the interior gravitational field
equations}

If the integrability condition given by Eq.~(\ref{int2}) is satisfied, the
general solution of Eq.~(\ref{3}) is given by
\begin{equation}
u(x)=\frac{1}{C_{A}+x}-\frac{1}{4}\frac{d}{dx}\ln \left( 1-2v\right) ,
\end{equation}%
where $C_{A}$ is an arbitrary integration constant. Therefore we obtain
\be
A(\xi ) =A_{0}\frac{\left( R_{c}^{2}\xi +C_{A}\right) }{\left[ 1-2v(\xi )\right]
^{1/4}}=
A_{0}R_{c}^{2}\frac{\left( \xi +C_{A}/R_{c}^{2}\right) }{V(\xi )},
\ee
where $A_{0}$ is an arbitrary integration constant. The density distribution
inside the star is found as
\bea\label{den}
\rho (\xi ) &=&\rho _{c}\left[ \frac{1-V^{4}(\xi )}{2\xi }+\frac{1}{3}\xi
\frac{d}{d\xi }\frac{1-V^{4}(\xi )}{\xi }\right] =\nonumber\\
 &&\frac{\rho _{c}}{6} \frac{1-V^{4}(\xi )}{\xi }\times
\left\{ 1+2\xi \frac{d}{d\xi }\ln \left[ 1-V^{4}\left( \xi \right) \right]
\right\} =\nonumber\\
&&\rho _{c}\rho _{0}(\xi ),
\eea
where we have denoted
\be
\rho _{0}(\xi ) =\frac{1-V^{4}(\xi )}{6\xi }\times
\left\{ 1+2\xi \frac{d}{d\xi }\ln \left[ 1-V^{4}\left( \xi \right) \right]
\right\} .
\ee
The mass distribution of the star is obtained as
\be
m(\xi ) =2\pi \rho _{c}R_{c}^{3}\int_{0}^{\xi }{\rho _{0}(\xi ')\sqrt{\xi '}%
d\xi '}=
2\pi \rho _{c}R_{c}^{3}m_{0}(\xi ),
\ee
where
\begin{equation}
m_{0}(\xi )=\int_{0}^{\xi }{\rho _{0}(\xi ' )\sqrt{\xi '}d\xi '}.
\end{equation}%
The metric tensor component $B$ can be written as
\begin{equation}
B(\xi )=1-\frac{3m_{0}(\xi )}{\sqrt{\xi }}.
\end{equation}%
And, finally, the pressure distribution in the star can be determined from
the equation
\bea \label{pree}
p(\xi )& =&\frac{2}{3}\rho _{c}c^{2}\Bigg\{\left[ \frac{1}{%
C_{A}/R_{c}^{2}+\xi }-\frac{d}{d\xi }\ln V(\xi )\right] \times \nonumber\\
&&\left[ 1-\frac{3m_{0}(\xi )}{\sqrt{\xi }}\right] -\frac{3}{4}\frac{%
m_{0}(\xi )}{\xi ^{3/2}}\Bigg\}=\frac{2}{3}\rho _{c}c^{2}p_0\left(\xi \right),\nonumber\\
\eea
where
\bea
p_0(\xi )&=&\left[ \frac{1}{%
C_{A}/R_{c}^{2}+\xi }-\frac{d}{d\xi }\ln V(\xi )\right] \times
\left[ 1-\frac{3m_{0}(\xi )}{\sqrt{\xi }}\right] -\nonumber\\
&&\frac{3}{4}\frac{%
m_{0}(\xi )}{\xi ^{3/2}}.
\eea
Thus the complete general solution of the gravitational field equations
describing the interior of a static isotropic general relativistic star
satisfying the Riccati integrability condition has been obtained.

\section{Physical properties of the solution}\label{sect4}

In order to be physically meaningful, the interior solution for static fluid
spheres of Einstein's gravitational field equations must satisfy some
general physical requirements. The following conditions have been generally
recognized to be crucial for isotropic fluid spheres \cite{Del}:

a) the density $\rho $ and pressure $p$ should be positive inside the star;

b) the gradients $d\rho /dr$ and $dp/dr$ should be negative;

c) inside the static configuration the speed of sound should be less than
the speed of light, i.e. $0\leq \left( 1/c^{2}\right) dp/d\rho \leq 1$;

d) the interior metric should be joined continuously with the exterior
Schwarzschild metric, that is $A^{2}(R)=1-2GM/c^2R$, where  $M$ is the
mass of the sphere as measured by its external gravitational field and $R$
is the boundary of the sphere;

e) the pressure $p$ must vanish at the boundary $r=R$ of the sphere.

\subsection{The astrophysical parameters of the star}

\begin{figure}
 \centering
 \includegraphics[scale=0.75]{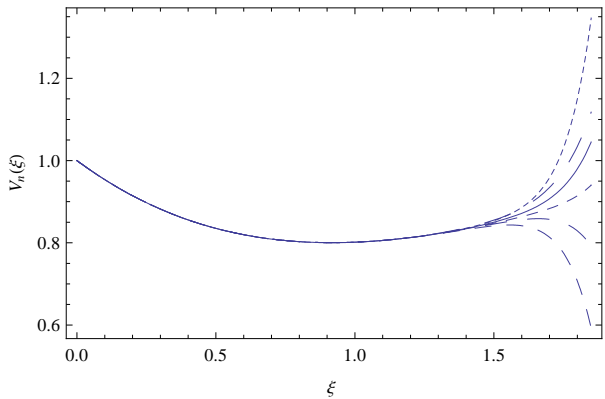}
 \caption{The variation of $V_n(\xi)=1-\xi/2+\sum_{k=2}^n{d_k\xi ^k}$ with respect to the dimensionless radial coordinate $\xi $ for $n=21$ (solid curve), $n=20$ (dotted curve), $n=19$ (dashed curve), $n=18$ (medium dashed curve), $n=17$ (long dashed curve), and $n=16$ (ultra-long-dashed curve, respectively).}
 \label{fig1}
\end{figure}

The general physical properties of the interior solution of the gravitational field equations obtained in the previous Section are determined by the behavior of the function $V\left(\xi \right)$, which at the center of the star satisfies the conditions $V(0)=1$, and $V'\left(0\right)=-1/2$. The variation of $V$ with respect to $\xi $ is represented, for several values of $n$, in Fig.~\ref{fig1}.

In the interval $\xi \in [0,1.5)$ all the solutions  $V_n\left(\xi\right)$, $n=16,17,18,19,20,21$ approximately coincide. However, for $\xi >1.5$, the qualitative and quantitative behavior of the function $V_n$ essentially depends on the value of $n$, and, consequently,  on the number of terms considered in the series expansion. In order to obtain definite numerical results, to maintain the {\it exact analytical} nature of the solution, and to avoid the excessive use of numerical methods, in the following we will restrict our study up to the case $n=21$ only.

\begin{figure}
 \centering
 \includegraphics[scale=0.75]{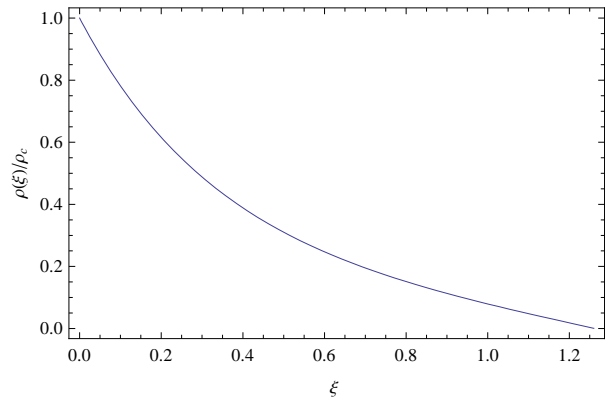}
 \caption{The variation of the ratio $\rho \left(\xi \right)/\rho _c $ of the density $\rho (\xi )$ at point $\xi $ and of the central density $\rho _c$ with respect to the dimensionless radial coordinate $\xi $.}
 \label{fig2}
\end{figure}

The function $V$ determines the behavior and properties of the density $\rho (\xi )$, whose variation with respect to $\xi $ in the stellar interior is represented in Fig.~\ref{fig2}.

The density is finite throughout the star, including the center $\xi =0$. For $\xi_ S=1.26$, the density vanishes, and $\rho (\xi )<0$ for $\xi >1.26$. Therefore we define the vacuum boundary of the star as the surface corresponding to $\xi =\xi _S$, with the property $\rho \left(\xi _S\right)=0$.
This choice determines the physical radius of the gaseous sphere as
\bea
R&=&R_c\sqrt{\xi _S}=\sqrt{\frac{3c^{2}}{4\pi G\rho _{c}}}\sqrt{\xi _S}=\nonumber\\
&&1.163\times
10^{6}\times \left( \frac{\rho _{c}}{3\times
10^{15}\;\mathrm{%
g/cm^{3}}}\right) ^{-1/2}\mathrm{cm}.
\eea

Using the density distribution inside the star given by Eq.~(\ref{den}), for the density gradient $d\rho (\xi )/d\xi $ we obtain the expression
\bea
\frac{d\rho (\xi )}{d\xi }&=&\frac{\rho _{c}}{6}\Bigg\{ 2\left[ 1-V^{4}(\xi )%
\right] \frac{d^{2}}{d\xi ^{2}}\ln \left[ 1-V^{4}\left( \xi \right) \right] -\nonumber\\
&&\frac{dV^{4}(\xi )}{d\xi }\left[ \frac{1}{\xi }+
2\frac{d}{d\xi }\ln \left[
1-V^{4}\left( \xi \right) \right] \right] +\nonumber\\
&&\frac{V^{4}(\xi )-1}{\xi ^{2}}\Bigg\} .  \label{den_grad}
\eea

\begin{figure}
 \centering
 \includegraphics[scale=0.75]{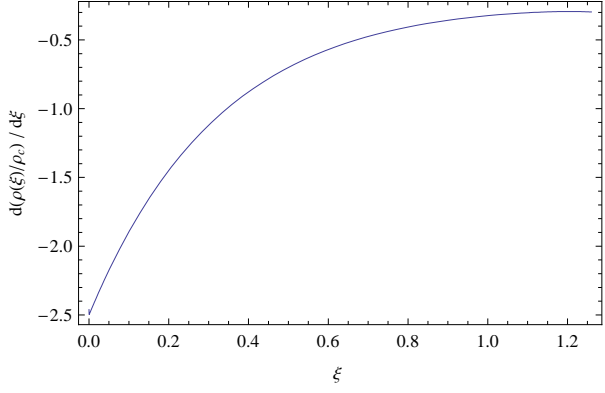}
 \caption{The variation of the density gradient $d\left[\rho (\xi )/\rho _c\right]/d\xi $ inside the star as a function of the dimensionless radial coordinate $\xi $.}
\label{figd}
\end{figure}

The variation of the density gradient with respect to $\xi $ is represented in Fig.~\ref{figd}. For all points in the range $\xi \in \left[0,\xi _S\right]$, the density gradient satisfies the condition $d\rho /d\xi <0$.

\begin{figure}
 \centering
 \includegraphics[scale=0.75]{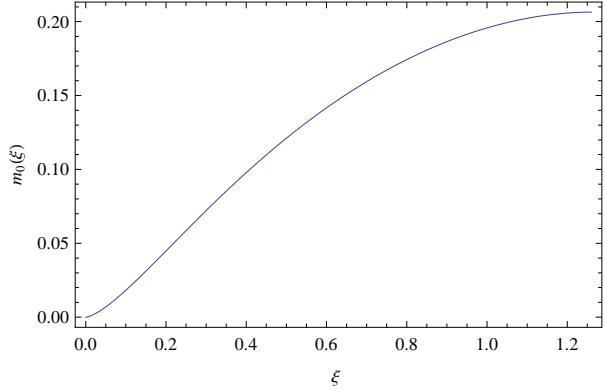}
 \caption{The variation of the dimensionless mass $m_0 (\xi )$ with respect to the dimensionless radial coordinate $\xi $.}
\label{fig3}
\end{figure}

The variation of the dimensionless mass function in the interval $\xi \in\left[0,\xi _S\right]$ is represented in Fig.~\ref{fig3}.
The mass profile inside the star is a monotonically increasing function. At the surface of the star $m_0$ reaches the  value $m_0\left(\xi _S\right)=0.206$. Therefore the total mass $M$ of the star is given by
\bea
M&=&2\pi \rho _c R_c^{3}m_0\left(\xi _S\right)=
\sqrt{\frac{27}{16 \pi }}\left(\frac{c^2}{G}\right)^{3/2}\frac{m_0\left(\xi _S\right)}{\sqrt{\rho _c}}=\nonumber\\
&&2.16\left(\frac{\rho _c}{3\times 10^{15}\;{\rm g/cm^3}}\right)^{-1/2}M_{\odot}.
\eea

\begin{figure}
 \centering
 \includegraphics[scale=0.75]{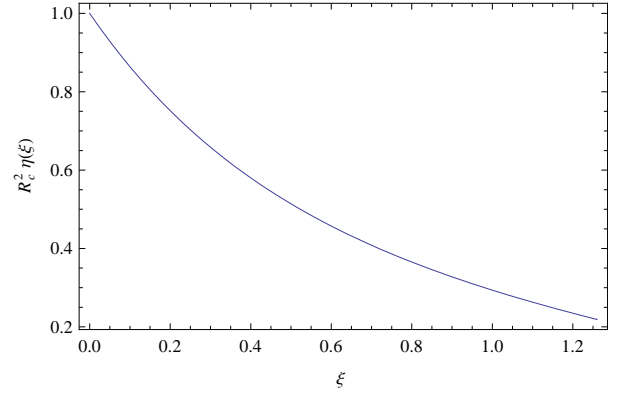}
 \caption{The variation of the mean density $R_c^2\eta (\xi )$ with respect to the dimensionless radial coordinate $\xi $.}
\label{eta}
\end{figure}

The variation of the mean density $\eta $ inside the star is represented in Fig~\ref{eta}.

 The mean density $\eta $ is a monotonically decreasing function of $\xi $, and on the surface of the star it takes the value $\eta \left(\xi _S\right)=0.218$. This allows us to define the mean density of the star as
 \be
 <\rho >=\frac{3M}{4\pi R^3}=\frac{3}{2}\rho _c\eta \left(\xi _S\right)=0.327\rho_c.
 \ee

 Next we consider the behavior of the pressure inside the star. By taking into account that $\lim_{\xi\rightarrow 0}V'(\xi)/V(\xi )=-1/2$, $3\lim_{\xi\rightarrow 0}m_0(\xi )/\sqrt{\xi }=0$, and $\lim_{\xi\rightarrow 0}m_0(\xi )/\xi ^{3/2}=2/3$, respectively,  from Eq.~(\ref{pree}) we obtain
\be
p(0)=p_c=\frac{2}{3}\rho _c c^2\frac{R_c^2}{C_A},
\ee
which allows us to express the arbitrary integration constant $C_A$ as a function of the central pressure and central density as
\be
\frac{C_A}{R_c^2}=\frac{2}{3}\frac{\rho _c c^2}{p_c}.
\ee

The ratio $C_A/R_c^2$ varies between $C_A/R_c^2=2/3\approx 0.66$, corresponding to the stiff causal equation of state of the form $p_c=\rho _cc^2$ at the center of the star, and $C_A/R_c^2=2$, corresponding to the extremely relativistic radiation type equation of state $p_c=\rho _cc^2/3$.  Hence the pressure distribution can be written as
\bea\label{press}
p(\xi)&=&\frac{2}{3}\rho _c c^2
\Bigg\{\Bigg[\frac{1}{(2/3)\rho _cc^2/p_c+\xi}-
\frac{V'(\xi)}{V(\xi)}\Bigg]\times \nonumber\\
&&\left[1-\frac{3m_0(\xi )}{\sqrt{\xi }}\right]-
\frac{3}{4}\frac{m_0(\xi)}{\xi ^{3/2}}\Bigg\}.
\eea

\begin{figure}
 \centering
 \includegraphics[scale=0.75]{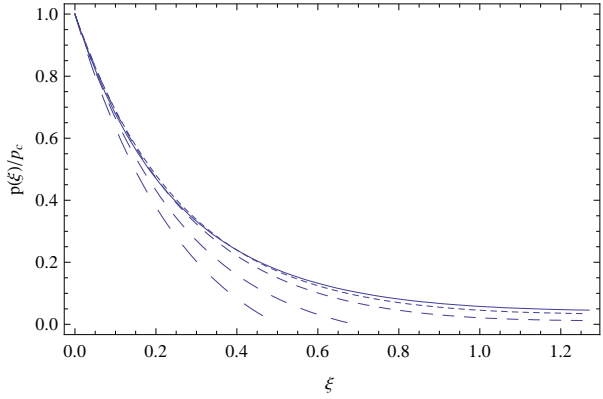}
 \caption{Variation of the ratio $p(\xi)/p_c$  with respect to the dimensionless radial coordinate $\xi $ for different equations of state of the stellar matter at the center of the star: $p_c=\rho _{c}c^2$ (solid curve), $p_c=(2/3)\rho _{c}c^2$ (dotted curve), $p_c=\rho _{c}c^2/3$ (short dashed curve), $p_c=\rho _{c}c^2/2$ (dashed curve), $p_c=\rho _{c}c^2/3$ (long dashed curve) and $p_c=\rho _{c}c^2/4$ (ultra-long dashed curve), respectively.  }
\label{pres}
\end{figure}

Therefore in order to construct specific stellar models the equation of state at the center of the star must also  be specified. The variation of the pressure as a function of the dimensionless radial coordinate $\xi $ is represented, for several equations of state of the matter at the center of the star, in Fig.~\ref{pres}.

The pressure tends to zero at the zero density limit $\xi_S=1.26$ in the case of the stiff ($p_c=\rho _cc^2$), or very similar ($p_c=2\rho _cc^2/3$) equations of state. In the case of the radiation equation of state the pressure reaches a zero value before the density. These results suggest that in the present model the equation of state of the matter at the center of the star is either the stiff causal equation of state, or a very similar one. Of course one could also consider physical models with a non-zero surface density, but vanishing surface pressure. For a stellar model with a radiation type central equation of state the pressure vanishes at $\xi _S=0.69$, which defines the radius of the star. The corresponding surface density is $\rho _S=\rho \left(\xi _S\right)=0.199\rho _c$, and $m_0\left(\xi _S\right)=0.157$. The physical radius and mass of this type of star are
\bea
R=8.607\times
10^{5}\times \left( \frac{\rho _{c}}{3\times
10^{15}\;\mathrm{%
g/cm^{3}}}\right) ^{-1/2}\mathrm{cm},
\eea
and
\bea
M=
1.646\times \left(\frac{\rho _c}{3\times 10^{15}\;{\rm g/cm^3}}\right)^{-1/2}M_{\odot},
\eea
respectively.

Using the pressure distribution inside the star as given by Eq.~(\ref{pree}), we obtain
the expression for the pressure gradient in the form
\begin{widetext}
\begin{eqnarray}
\frac{dp(\xi )}{d\xi }&=&2\rho _{c}c^{2}\Bigg\{ \left[ \frac{d}{d\xi }\ln
V\left( \xi \right) -\frac{1}{C_{A}/R_{c}^{2}+\xi }\right]
\left[ \rho
_{0}(\xi ){-}\frac{m_{0}(\xi )}{2\xi ^{3/2}}\right] -
\left[ \frac{1}{3}-\frac{m_{0}(\xi )}{\sqrt{\xi }}\right]\times \nonumber\\
&&\left[ \frac{1}{%
\left( C_{A}/R_{c}^{2}+\xi \right) ^{2}}+\frac{d^{2}}{d\xi ^{2}}\ln V\left(
\xi \right) \right] -
\frac{1}{4\xi }\left[ \rho _{0}(\xi ){-}\frac{3m_{0}(\xi )}{2\xi ^{3/2}}%
\right] \Bigg\} .
\end{eqnarray}
\end{widetext}

\begin{figure}
 \centering
 \includegraphics[scale=0.75]{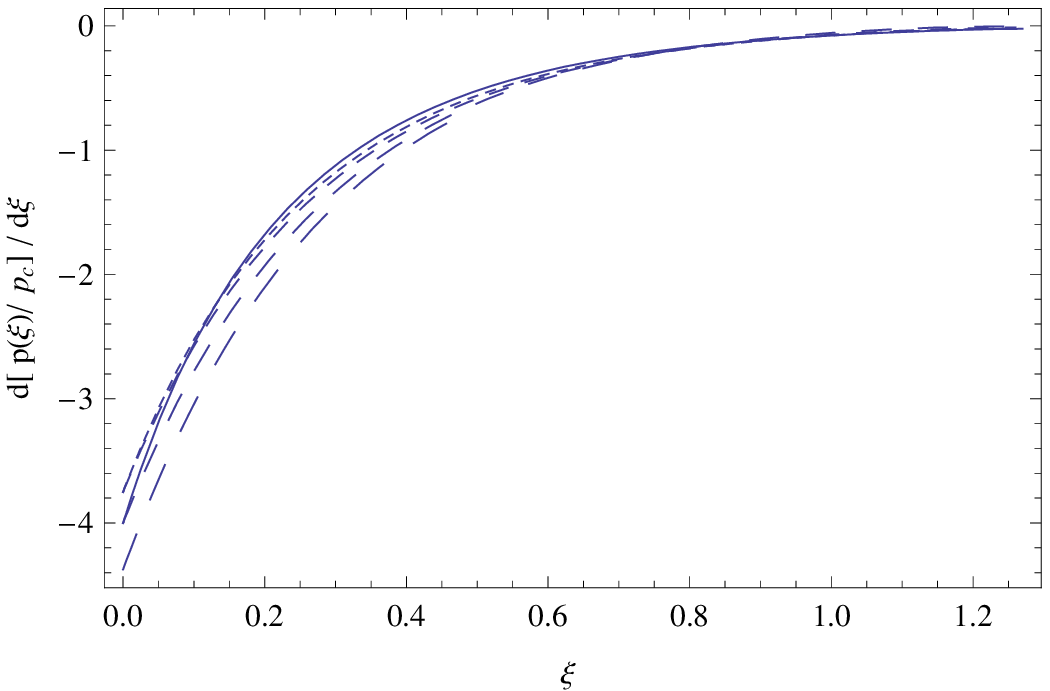}
 \caption{Variation of the pressure gradient $d\left[p(\xi)/p_c\right]/d\xi $  with respect to the dimensionless radial coordinate $\xi $ for different equations of state of the stellar matter at the center of the star: $p_c=\rho _{c}c^2$ (solid curve), $p_c=(2/3)\rho _{c}c^2$ (dotted curve), $p_c=\rho _{c}c^2/3$ (short dashed curve), $p_c=\rho _{c}c^2/2$ (dashed curve), $p_c=\rho _{c}c^2/3$ (long dashed curve) and $p_c=\rho _{c}c^2/4$ (ultra-long dashed curve), respectively. }
\label{dpres}
\end{figure}

The variation of $dp(\xi )/d\xi $ inside the star is represented in Fig.~\ref{dpres}.

The pressure gradient satisfies the condition $dp(\xi )/d\xi<0$ for all $\xi \in [0,\xi _S]$. In view of the density gradient and of the pressure gradient, it is easy to
obtain the speed of sound $c_s^2=dp/d\rho $ inside the star as
\begin{widetext}
\begin{eqnarray}\label{speed}
&&\frac{c_s^2}{c^{2}} =12
\frac{\left( \frac{d}{d\xi }\ln V -\frac{1}{%
C_{A}/R_{c}^{2}+\xi }\right) \left( \rho _{0}{-}\frac{m_{0}}{%
2\xi ^{3/2}}\right) -\frac{B }{3}\left[ \frac{1}{\left(
C_{A}/R_{c}^{2}+\xi \right) ^{2}}+\frac{d^{2}}{d\xi ^{2}}\ln V \right] -\frac{1}{4\xi }\left[ \rho _{0}{-}\frac{3m_{0}}{%
2\xi ^{3/2}}\right] }{2\left[ 1-V^{4}\right] \frac{d^{2}}{d\xi ^{2}}%
\ln \left[ 1-V^{4} \right] -\frac{dV^{4}}{d\xi }%
\left[ \frac{1}{\xi }+2\frac{d}{d\xi }\ln \left[ 1-V^{4} %
\right] \right] +\frac{V^{4}-1}{\xi ^{2}}}. \nonumber\\
\end{eqnarray}
\end{widetext}

\begin{figure}
 \centering
 \includegraphics[scale=0.75]{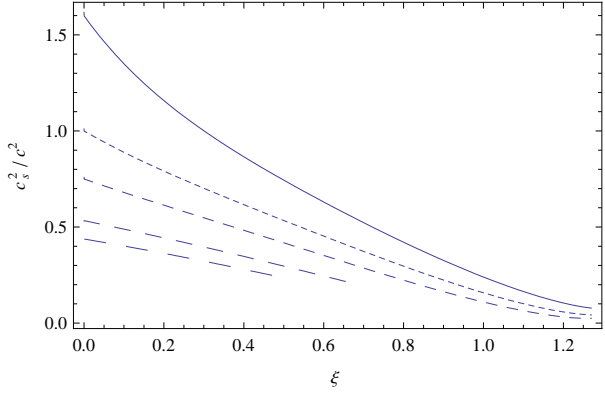}
 \caption{Variation of the ratio $c_s^2/c^2$  with respect to the dimensionless radial coordinate $\xi $ for different equations of state of the stellar matter at the center of the star: $p_c=\rho _{c}c^2$ (solid curve), $p_c=(2/3)\rho _{c}c^2$ (dotted curve), $p_c=\rho _{c}c^2/3$ (short dashed curve), $p_c=\rho _{c}c^2/2$ (dashed curve), $p_c=\rho _{c}c^2/3$ (long dashed curve) and $p_c=\rho _{c}c^2/4$ (ultra-long dashed curve), respectively. }
\label{cs}
\end{figure}

The ratio of the speed of sound and of the speed of light inside the stellar interior is presented for two equations of state at the center in Fig.~\ref{cs}.

As expected, the speed of sound decreases outwards in the fluid sphere. For all points inside the star $c_s$ satisfies the condition $c_s\leq c$ for $p_c\leq 2\rho _cc^2/3$. However, we have to point out that the definition of the speed of sound we have used is valid only for isentropic fluids, that is, for fluids in which the entropy per baryon is constant throughout the star. In general the definition of the speed of sound is $c_s^2=\left(\partial p/\partial \rho \right)_s$, where $s$ is the entropy per baryon \cite{Knut}.

\begin{figure}
 \centering
 \includegraphics[width=80mm, height=50mm]{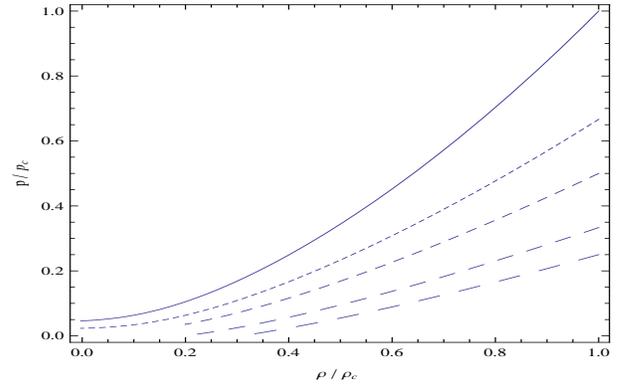}
 \caption{ The equation of state $p/p_c=f\left(\rho /\rho _c\right)$ of the matter inside the star for different equations of state of the stellar matter at the center of the star: $p_c=\rho _{c}c^2$ (solid curve), $p_c=(2/3)\rho _{c}c^2$ (dotted curve), $p_c=\rho _{c}c^2/3$ (short dashed curve), $p_c=\rho _{c}c^2/2$ (dashed curve), $p_c=\rho _{c}c^2/3$ (long dashed curve) and $p_c=\rho _{c}c^2/4$ (ultra-long dashed curve), respectively.  }
\label{eos}
\end{figure}

An important physical characteristic of general relativistic objects,  the equation of state of the dense matter inside the star is represented in Fig.~\ref{eos}. The equation of state of the matter cannot be obtained in an analytical form.

Finally, we consider the matching of the metric coefficient $A$ on the stellar surface with a vacuum boundary. First of all we  redefine the arbitrary integration constant $A_0$ so that $A_0=\alpha /R_c^2$, where $\alpha $ is a dimensionless constant. On the surface of the star, where $\xi _S=1.26$, $V(\xi _S )=0.818$, and the matching condition for $A$ becomes
\be
\alpha \frac{\xi _S+(2/3)\rho _cc^2/p _c}{V(\xi _S)}=\sqrt{1-\frac{3m_0(\xi _S)}{\sqrt{\xi _S}}},
\ee
which gives for $\alpha $ the expression
\be
\alpha =\frac{0.739}{1.26+(2/3)\rho _cc^2/p _c}.
\ee

\subsection{Dynamical stability properties of the stellar model}

In the following we study the stability of the stellar model with respect to
the infinitesimal radial adiabatic perturbations by using the approach
introduced in \cite{Chand}, and further developed and applied to concrete
astrophysical cases in \cite{Bar} and \cite{Knut}, respectively. A normal
mode of radial oscillation for an equilibrium configuration  is stable if $%
\delta r=\zeta (r)\exp \left( i\sigma x^0\right) $, where $x^0=ct$, is stable (periodic
oscillation) if its frequency $\sigma $ is real, and unstable (exponential
growth), if $\sigma $ is imaginary. Here $\zeta (r)$ is a chosen
trial function that must satisfy some appropriate boundary conditions.

By assuming that all the perturbations have a dependence on $x^{0}$ of
the form $\exp (i\sigma x^{0})$, from the linearized Einstein field
equations we obtain the Sturm-Liouville eigenvalue equation for the
eigenmodes  as
\begin{equation}
\frac{d}{dr}\left( \Pi \frac{d\zeta _{j}}{dr}\right) +\left( \Theta +\sigma
_{j}^{2}W\right) \zeta _{j}=0,\ j=1,2,...,n,  \label{eig}
\end{equation}
where we have redefined the ``Lagrangian displacement'' $\chi $ as $\chi
\rightarrow r^{-2}A(r)\zeta $ (corresponding to a Lagrangian displacement of
the radial coordinate of the form $\delta r(x^{0},r)=r^{-2}A(r)\zeta \exp
(i\sigma x^{0})$, and we have denoted
\begin{equation}
\Pi =\frac{\gamma p}{r^{2}%
}\frac{A^{3}}{\sqrt{B}},
\end{equation}
\begin{equation}
\Theta =\frac{\Pi }{\gamma p}\left[ \frac{p^{\prime 2}}{\rho c^{2}+p}-\frac{%
4p^{\prime }}{r}-\frac{8\pi G}{c^{4}}p(\rho c^{2}+p)\frac{1}{B}\right] ,
\end{equation}
and
\begin{equation}
W=\frac{\rho c^{2}+p}{r^{2}}\frac{A}{B^{3/2}},
\end{equation}
respectively. The adiabatic index $\gamma $ is defined as
\begin{equation}
\gamma =\frac{\rho c^{2}+p}{p}\frac{1}{c^{2}}\frac{dp}{d\rho }.
\end{equation}

The boundary conditions for $\zeta (r)$ are that $\zeta (r)/r^{3}$ is finite
or zero as $r\rightarrow 0$ and that the Lagrangian variation of the
pressure $\Delta p=-\left( \gamma pA/r^{2}\right) d\zeta /dr$ vanishes at
the surface of the star.

The eigenvalue problem formulated in Eq.~(\ref{eig}) can be re-expressed in
a well-known variational form: the extremal values of the quantity
\begin{equation}
\sigma ^{2}=\frac{c^2\int_{0}^{R}\left[ \Pi \left( \frac{d\zeta }{dr}\right)
^{2}-\Theta \zeta ^{2}\right] dr}{\int_{0}^{R}W\zeta ^{2}dr},  \label{var}
\end{equation}
are the eigenvalues $\sigma _{j}^{2}$ of Eq. (\ref{eig}) and the functions $%
\zeta (r)$, which give the extremal values, are the corresponding
eigenfunctions. A sufficient condition for the dynamical instability of a
mass is that the right hand side of Eq. (\ref{var}) vanishes for some chosen
trial function $\zeta $ which satisfies the boundary conditions \cite{Chand, Bar}. For all stellar models of physical interest the frequency
spectrum of the normal radial modes is discrete; the $n$-th normal mode has $%
n$ nodes between the center and the surface of the star. The normal mode
eigenfunctions are orthogonal with respect to the weight function $W(r)$: $%
\int_{0}^{R}W\zeta _{j}\zeta _{k}dr=0$, if $j\neq k$. By using the explicit
forms of the metric functions we obtain
\begin{equation}
\Pi =\frac{\rho _{c}c^{2}}{R_{c}^{2}}\alpha ^{3}\gamma p_{0}\frac{\left[ \xi
+\left( 2/3\right) \rho _{c}c^{2}/p_{c}\right] ^{3}}{\xi V^{3}\sqrt{1-3m_{0}/%
\sqrt{\xi }}}=\frac{\rho _{c}c^{2}}{R_{c}^{2}}\alpha ^{3}\Pi _{0}\left( \xi
\right) ,
\end{equation}
\bea
\Theta &=&\frac{\rho _{c}c^{2}}{R_{c}^{4}}\alpha ^{3}\frac{\left[ \xi
+\left( 2/3\right) \rho _{c}c^{2}/p_{c}\right] ^{3}}{\xi V^{3}\sqrt{1-3m_{0}/%
\sqrt{\xi }}}\times \nonumber\\
&&\left[ \frac{4\xi \left(
dp_{0}/d\xi \right) ^{2}}{\rho _{0}+p_{0}}-8\frac{dp_{0}}{d\xi }%
-\frac{6p_{0}\left( \rho _{0}+p_{0}\right) }{ \sqrt{1-3m_{0}/\sqrt{\xi }}}%
\right] =\nonumber\\
&&\frac{\rho _{c}c^{2}}{R_{c}^{4}}\alpha ^{3}\Theta _{0}\left( \xi \right) ,
\eea
and
\be
W=\frac{\rho _{c}c^{2}}{R_{c}^{2}}\alpha \left(\rho _0+p_0\right)
\frac{\xi +\left( 2/3\right) \rho
_{c}c^{2}/p_{c}}{\xi V\left( 1-3m_{0}/\sqrt{\xi }\right) ^{3/2}}=
\frac{\rho
_{c}c^{2}}{R_{c}^{2}}\alpha W_{0}\left( \xi \right) ,
\ee
respectively.  For the oscillation frequency we find
\begin{equation}
\sigma ^{2}=\alpha ^{2}\frac{c^{2}}{R_{c}^{2}}\frac{\int_{0}^{\xi
_{S}}\left( 4\sqrt{\xi }\Pi _{0}\left( d\zeta /d\xi \right) ^{2}-\Theta _{0}\zeta
^{2}/\sqrt{\xi }\right) d\xi }{\int_{0}^{\xi _{S}}\left( W_{0}\zeta ^{2}/%
\sqrt{\xi }\right) d\xi }.
\end{equation}

\begin{figure}
 \centering
\includegraphics[scale=0.75]{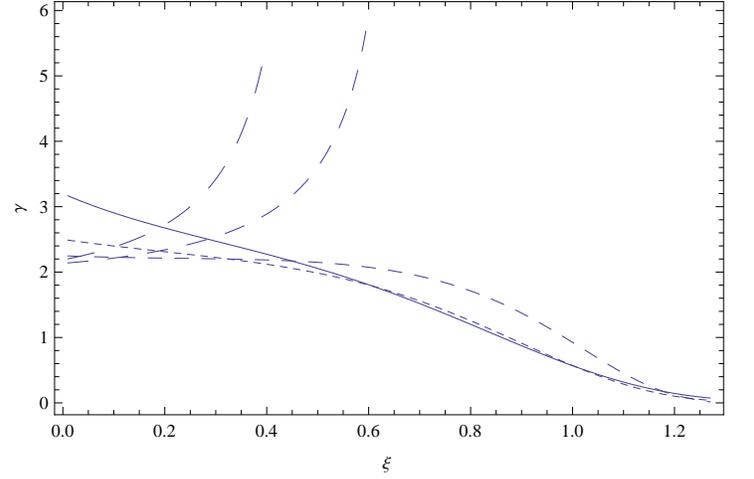}
\caption{Variation of the adiabatic index $\gamma $  with respect to the dimensionless radial coordinate $\xi $ for different equations of state of the stellar matter at the center of the star: $p_c=\rho _{c}c^2$ (solid curve), $p_c=(2/3)\rho _{c}c^2$ (dotted curve), $p_c=\rho _{c}c^2/3$ (short dashed curve), $p_c=\rho _{c}c^2/2$ (dashed curve), $p_c=\rho _{c}c^2/3$ (long dashed curve) and $p_c=\rho _{c}c^2/4$ (ultra-long dashed curve), respectively.  }
\label{gamma}
\end{figure}

The variation of the adiabatic index $\gamma $ inside the star is represented in Fig.~\ref{gamma}.

Depending on the equation of state at the center, the adiabatic index shows a large variety of behaviors. For $1/2\leq p_c/\rho _cc^2\leq 1$, at the center of the star the adiabatic index has values in the range of $2.2\leq \gamma (0) \leq 3.2$, and then $\gamma $ monotonically  decreases to zero at the surface of the star. For $p_c/\rho _cc^2\leq 1/3$, the central value of the adiabatic index is in the range $2.2-2.5$, and then it rapidly increases to values of the order of 5 at the surface of the star. To find a physical interpretation of the behavior of $\gamma $ we adopt a simplified physical model in which we assume that at each point the matter obeys the perfect gas equation of state $p(r)=\left(k_B/\mu m_H\right)\rho (r)T(r)$, where $k_B$ is Boltzmann's constant, $\mu $ is the atomic weight of the matter, $m_H$ is the proton mass, and $T(r)$ is the local temperature. By using the definition of $\gamma $ we obtain the temperature gradient inside the star as
\be
\frac{dT}{d\rho }=\frac{\mu m_H}{k_B}\left(\gamma -1-\frac{c_s^2}{c^2}\right).
\ee

For $p_c/\rho _cc^2\leq 1/3$, $dT/d\rho >0,\forall \xi \in \left(0,\xi _S\right)$. Therefore the temperature is monotonically increasing with the density inside the star, reaching its maximum value at the center. On the other hand for $1/2\leq p_c/\rho _cc^2\leq 1$, the temperature gradient changes its sign inside the star, so that in some regions there is a transition from a state of matter with $dT/d\rho >0$ to a phase with $dT/d\rho <0$, which may suggest the presence of some thermal instabilities inside the star.

The values of the pulsation frequency are presented, for different equations of state at the center, and for different trial functions $\zeta $, in Table~\ref{t1}. As one can see from the Table, for all chosen trial functions the oscillations frequencies are positive, which shows that the stellar model is stable against radial adiabatic infinitesimal perturbations for all considered central equations of state.

\begin{table}[h]
\begin{tabular}{|l|l|l|}
\hline
$p_{c}/\rho _{c}c^{2}$ & $\zeta $ & $R_{c}^{2}\sigma ^{2}/c^{2}$ \\
\hline
1 & $\xi ^{3}$ & 0.167 \\
1 & $e^{\xi /2}\xi ^{3}$ & 0.108 \\
1 & $e^{-\xi /2}\xi ^{4}$ & 0.245 \\
1 & $e^{-\xi /2}\xi ^{7/2}$ & 0.174 \\
\hline
2/3 & $\xi ^{3}$ & 0.082 \\
2/3 & $e^{\xi /2}\xi ^{3}$ & 0.114 \\
2/3 & $e^{-\xi /2}\xi ^{4}$ & 0.122 \\
2/3 & $e^{-\xi /2}\xi ^{7/2}$ & 0.088 \\
\hline
1/2 & $\xi ^{3}$ & 0.047 \\
1/2  & $e^{\xi /2}\xi ^{3}$ & 0.063 \\
1/2 & $e^{-\xi /2}\xi ^{4}$ & 0.070 \\
1/2 & $e^{-\xi /2}\xi ^{7/2}$ & 0.051 \\
\hline
1/3 & $\xi ^{3}$ & 0.143 \\
1/3 & $e^{\xi /2}\xi ^{3}$ & 0.174 \\
1/3 & $e^{-\xi /2}\xi ^{4}$ & 0.224 \\
1/3 & $e^{-\xi /2}\xi ^{7/2}$ & 0.166 \\
\hline
1/4 & $\xi ^{3}$ & 0.167 \\
1/4 & $e^{\xi /2}\xi ^{3}$ & 0.191 \\
1/4 & $e^{-\xi /2}\xi ^{4}$ & 0.266 \\
1/4 & $e^{-\xi /2}\xi ^{7/2}$ & 0.201 \\
\hline
\end{tabular}
\caption{ Oscillation frequencies of the stellar model for different equations of state at the center of the star, and for different trial functions $\zeta $.}\label{t1}
\end{table}

\section{Discussions and final remarks}\label{sect5}

In the present paper we have obtained an exact analytical solution of the gravitational field equations describing the interior of massive general relativistic fluid spheres. The solution was obtained by using an integrability condition of the Riccati equation, which is equivalent with the condition of the pressure isotropy inside the star. The integrability condition reduces to a second order non-linear differential equation, which can be solved exactly in terms of an infinite power series. This generates an exact, non-singular solution of the interior field equations, which gives a full physical description of the physical parameters of the system. We have analyzed in detail the physical properties of the star corresponding to this exact solution. Its physical properties critically depend on the equation of state of the matter at the center of the star. Since the equation of state of the matter at high densities is unknown, we have restricted our analysis to the case in which the central matter is described by a simple linear barotropic equation of state $p_c=\beta \rho _cc^2$, and we have analyzed in detail the cases $\beta =1,2/3,1/2,1/3$ and $1/4$, respectively. In particular, the speed of sound is smaller than the isentropic speed of light for $\beta \leq 2/3$, with the isentropic speed of sound exceeding the speed of light for $2/3\leq \beta \leq 1$. However, all analyzed models are stable with respect to the radial adiabatic perturbations, thus showing that for the present case the possible violation of the causality condition does not automatically lead to a physical instability of the stellar object. In fact all considered models are stable with respect to infinitesimal radial perturbations. On the other hand, at high densities new phases of matter can appear in the high density core of compact objects, like quark, kaon, hyperon or different types of Bose-Einstein condensates \cite{Cha}. Hence a more realistic choice of the equation of state at the center of the star may lead to different results for the causality condition, as well as for the behavior of the adiabatic index $\gamma $.

From the study of the astrophysical properties of pulsars presently there is conclusive observational evidence  for the existence of neutron stars with
masses significantly greater than $1.5M_{\odot}$ \cite{Lat}. By using the Shapiro time delay to measure the inclination  the mass of the pulsar PSR J1614-223048 was recently determined with a high accuracy to be  $1.97\pm 0.04 M_{\odot}$ \cite{Dem}. Moreover, a relatively high number of X-ray binaries seem to contain high-mass neutron
stars: about $1.9M_{\odot}$ in the case of Vela X-1 and $2.4M_{\odot}$ in the case of 4U 1700-377 \cite{Lat}. Even more interesting, and intriguing, is the case of the  black widow pulsar B1957+20, with a best mass estimate of
about $2.4M_{\odot}$ \cite{black}. For this system  both pulsar timing and optical light curve information are available. It consists of  pulsar with a 1.6 ms period in a nearly circular 9.17 hours orbit around an
extremely low mass companion, of mass $M_c\approx  0.03M_{\odot}$. The irradiation of the companion star by the pulsar strongly heats the cosmic environment to the point of ablation, creating  a comet-like tail, and a large cloud of plasma, with the plasma cloud
determining  the observed eclipsing. Therefore the pulsar is literally consuming its companion, and
hence the name "black widow" for this cosmic object. During the long time astrophysical evolution the mass of the companion star has been reduced  to a small fraction of its
 initial mass \cite{Lat}. On the other hand, the theoretical analysis of the structure of the pulsar shows that a measured mass of $2.4 M_{\odot}$ would be incompatible with hybrid star models containing {\it significant} proportions of exotic matter in the form of hyperons, some forms of Bose condensates, or quark matter \cite{Lat}.

However, we would like to point out that the mass and radius of the $2-2.4M_{\odot}$ neutron stars perfectly fit the expected
properties of a star in the present general relativistic stellar model. For a central density of the order of $\rho _c \sim 3\times 10^{15}$ g/cm$^3$, the mass of a typical general relativistic  star described by the  solution of the field equations
 is of the order of two solar masses, with a radius of around 11
km. Therefore, we suggest that the recently observed $2-2.4M_{\odot}$ mass neutron stars could be  a typical
example of a general relativistic star described by the static model obtained in the present paper.

\section*{Acknowledgments}

We would like to thank to the anonymous referee for comments and suggestions that helped us to improve our manuscript.

\end{document}